\documentclass[prb]{revtex4}
\usepackage{graphicx}
\begin{document}
\title{The role of the alloy structure in the magnetic behavior of granular systems}
\author{C. S. M. Bastos, M. Bahiana, W. C. Nunes and M. A. Novak}
\affiliation{Instituto de F\'{\i}sica, Universidade Federal do Rio de Janeiro\\
Caixa Postal 68528, Rio de Janeiro, RJ, Brazil, 21945-970}
\author{D. Altbir}
\affiliation{Departamento de F\'{\i}sica, Universidad de Santiago de Chile\\
Casilla 307, Santiago 2, Chile}
\author{P. Vargas}
\affiliation{Departamento de F\'{\i}sica, Universidad T\'ecnica Federico Santa Mar\'{\i}a\\
Casilla 110-V, Valpara\'{\i}so, Chile}
\author{M. Knobel}
\affiliation{Instituto de F\'{\i}sica ``Gleb Wataghin'', Universidade
Estadual de Campinas (UNICAMP)\\
CP 6165, 13083-970, Campinas, S\~ao Paulo, Brazil}
\date{\today}
\begin{abstract}
The effect of grain size, easy magnetization axis and anisotropy constant
distributions in the irreversible magnetic behavior of granular alloys is considered. A simulated granular
alloy is used to provide a realistic grain structure for the Monte Carlo simulation
of the ZFC-FC curves. The effect of annealing and external field is also studied.
The simulation curves are in good agreement with
the FC and ZFC magnetization curves measured on melt spun
Cu-Co ribbons.
\end{abstract}
\maketitle

\section{Introduction }\label{sec:Intro}

Besides the importance of studying still open problems on basic magnetism,
nanocrystalline systems attract more and more interest due to their applications on
chemical catalysis and magnetic recording \cite{recording}. For the latter, the progressive increase of
recording density has lead to the production of materials with smaller single domain particles.
This requirement has the serious drawback that the effective magnetic moment of the particles
suffer strong instabilities of thermal origin, so called superparamagnetic limit. Also, when
one deals with the nanometer scale, the magnetic systems are not easily reproduced and
characterized, introducing additional difficulties for experimental studies.  Experimental
and theoretical results over the past years show that there are clearly many factors that
can influence the magnetic and magnetotransport behavior of these systems, namely: The
distribution of grain sizes, the average size and shape of the grains, the magnetic
anisotropy of the individual grains and magnetic interactions among the
nanometer-sized crystallites.

The effect of some of these parameters on the magnetization and
magnetoresistance has been partially investigated by several authors.
El-Hilo {\it et al.}\cite{el-hilo} have used Monte Carlo simulations for
determining the magnetoresistance dependence on the mean intergranular
distance, or rather, the particle concentration, using a simple
expression previously obtained by Gittleman {\it et al.}\cite{gittleman}.
The influence of the log-normal distribution of magnetic moments and the
ratio of the boundary to the volume scattering cross sections on the
magnetization and magnetoresistance has been examined by Ferrari {\it et al.}
\cite{ferrari}.  Hickey {\it et al.}\cite{hickey} and Wiser\cite{wiser}
have developed a phenomenological model which explains the almost linear
variation of the magnetoresistance with the magnetization at low
temperatures as a possible consequence of existing correlations between
blocked and superparamagnetic particles. The blocking effect on the
magnetization curves has been simulated by a Monte Carlo method in a
paper of Dimitrov and Wysin \cite{dimitrov}. Allia {\it et al.}
\cite{allia} have proposed analytical models that take explicitly into
account the correlation arising from the dipolar interactions on nearly
superparamagnetic systems. Also for interacting systems, Chantrell {\it et
al.}\cite{Chantrell} calculated the susceptibillity and Field-Cooling (FC)
and Zero Field-Cooled (ZFC) curves for
superparamagnetic particles, and Pike {\it et al.}\cite{Pike} investigated
the role of magnetic interactions on low temperature saturation remanence
of fine magnetic particles. However, most of these models assume
non-realistic size distributions of the grains, and  constant
anisotropy, as if all grains had the same shape. Both parameters
originate effects that certainly are interesting to elucidate.

In order to investigate the effect of different structural and magnetic factors on the
magnetization properties of granular alloys we have simulated, by means of a  standard
Monte Carlo method, one of the most common characterization techniques, the FC-ZFC curves.
The samples were simulated by a Cell Dynamical System  model which gives the realistic
grain sizes and shapes distributions and, with the aid of experimental results,
these quantities can be assigned to magnetization and anisotropy for each grain.
After these quantities are known, the magnetic properties of the system are studied
as a function of temperature, applied field and thermal treatment of the sample.

Comparisons between the realistic distribution and others are made to check the
influence of grain sizes and shape distributions on ZFC-FC curves. An interesting
correspondence is made between annealing and time iteration in CDS. The applied field
effects on blocking temperatures and magnetization values for ZFC-FC curves are studied
and agree with experiments. Our results are compared with experimental
results also included is this work.

\section{Simulation of the Granular Alloy }\label{sect:alloy}

Granular materials correspond to metastable states resulting from the slowing down of the
segregation process in binary off-critical mixtures \cite{dg8}. In the case of metal alloys,
a quench to the metastable region of the phase diagram generates a virtually permanent
granular state at room temperature. Since the grain structure is not an equilibrium state, it
cannot be derived from a minimization
procedure. There are several statistically equivalent granular structures corresponding
to different paths in phase space and one must go through the process of phase separation in order to
reach one of these intermediate states and obtain
realistic simulations of granular alloys.

The traditional approach to this problem is the Cahn-Hilliard equation \cite{allencahn1}
\begin{equation}\label{CHC}
    \frac{\partial\psi(\mathbf{r},t)}{\partial t} = L \nabla^2 \frac{\delta F [\psi
    (\mathbf{r},t)]}{\delta \psi(\mathbf{r},t)} \,\, ,
\end{equation}
where $\psi(\mathbf{r},t)$ is a conserved order parameter (in our case, the
difference between Co and Cu concentrations), $L$ is a phenomenological
parameter related to the mobility, and $F[\psi(\mathbf{r},t)]$ is the coarse-grained Landau
free-energy functional:
\begin{equation}\label{fe}
    F[\psi(\mathbf{r},t)] = \int d\mathbf{r} \Bigg[ \frac{1}{2}(\nabla \psi)^2 -
    \frac{\tau}{2} \psi^2 + \frac{g}{4} \psi^4 \Bigg] \,\, ,
\end{equation}
with $\tau$ and $g$ positive phenomenological parameters.  In principle, the solution of
equation (\ref{CHC}) yields the knowledge of the whole phase separation process.  The problem
is, of course, that (\ref{CHC}) does not have an analytical solution and it is very hard to
solve numerically. Almost all efforts in this sense were able to describe the very early
stages of the process only.  Other common approaches are Monte Carlo simulations with Kawasaki
exchange dynamics \cite{binder1,knobel1} and Cell Dynamical System (CDS) modeling
\cite{oopu1,oopu2,ooyeung1,shinooo2,ooba1,mogo1,masscross,swmm1}. Since Monte Carlo simulations are computationally intensive and
CDS has been successfully used to model this type of problems, we have chosen the latter.

The basic point of CDS modeling is the discreteness of space and time. The point of view is
equivalent to the one in the Cahn-Hilliard equation, in the sense that a coarse grained description
based on densities is used, and the parameters are phenomenological.
The stability of the dynamics and the computational efficiency allows the achievement
of the asymptotic regime even for reasonably large systems. The  disadvantage of this approach is that the
arbitrary definition of parameters does not yield absolute information about the system, although this can be done in some
cases by recovering the correct time and space scales \cite{swmm1}.

The simulation, in this case,  mimics an experiment in which a binary alloy is initially prepared at a
temperature above the phase separation critical temperature, with given amounts of each
component, and then quenched to a region of the phase diagram in which the homogeneous phase
is metastable.  The system is represented by a $d$-dimensional lattice and its
configurations at time $t$ and $t+1$ are directly related by a map. Here we have considered only
$d=2$, since we believe that most of the properties we seek will be manifested in this
dimension, so the computational effort involved in a $3d$ simulation would not
improve significantly our results.

 We assume
that the dynamics of each lattice cell is governed by a local relaxational mechanism driven
by a suitable map {\it f}. The exact form of the map is not important, as long as it has the
correct flow \cite{oopu1,oopu2,puoo1}. For the segregation of binary mixtures we seek a map
with one unstable fixed point at the origin and two hyperbolically stable fixed points at
symmetrical positions. The stable fixed points correspond to the segregated phases rich in Co
and Cu, and the unstable fixed point, to the homogeneous phase.

The single cell dynamics is described by
\begin{equation}
    \psi(t+1, n) = f[\psi(t,n)] \,\, ,
\end{equation}
where $\psi(t,n)$ is the value of the order parameter in the cell $n$ at time
$t$. The addition of a diffusional coupling to its neighborhood leads to a non conservative
dynamics of the form
\begin{eqnarray}
    \psi(t+1, n) &=& f[\psi(t, n)] + D [ \langle\langle\psi(t,n)\rangle\rangle -\psi(t, n)] \\
     &=&\psi(t,n)+I(t,n)\,\, ,
\end{eqnarray}
where $I(t,n)=f[\psi(t,n)]-\psi(n,t)+
D\left[\langle\langle \psi(n,t)\rangle\rangle-\psi(n,t)\right]$ is the increment in order parameter after
one iteration, and
$D$ is a positive parameter proportional to the phenomenological diffusion
constant. $\langle\langle * \rangle\rangle - \; *$ is the isotropic discrete Laplacian.  We
use the following definition of spatial average $\langle\langle * \rangle\rangle$ on the two-dimensional
square lattice:
\begin{equation}
    \langle\langle\psi(t, n)\rangle\rangle \; =
\; \frac{1}{6} \sum_{nn} \; \psi + \frac{1}{12} \sum_{nnn} \; \psi   \,\, ,
\end{equation}
where the sums are over nearest-neighbors and next-nearest-neighbors cells
respectively. The conservation of order parameter may be locally imposed by the subtraction
of the average gain in order parameter in the neighborhood of each cell.
The conserved order parameter CDS model then reads
\begin{equation}
    \psi(t+1, n) = \psi(t,n)+I(t,n)-\langle\langle I(t,n)\rangle\rangle \,\, .
\label{cdsmodel}
\end{equation}
\subsection{Simulations conditions}
\begin{figure}
\includegraphics{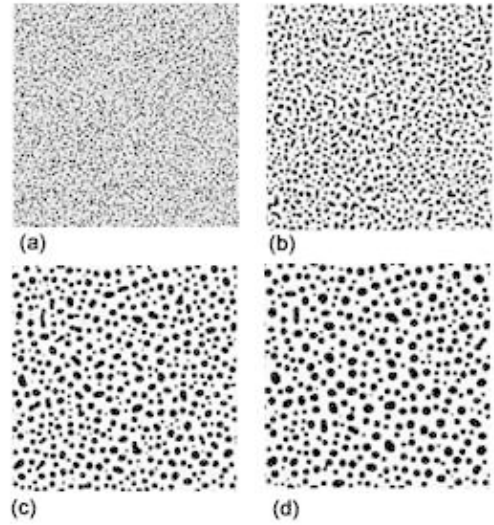}
 \caption{Segregation patterns for a Cu$_{90}$Co$_{10}$ alloy after (a) 100, (b)
1000, (c) 5000, and (d) 10000 iterations of Eq. (\ref{cdsmodel}). Darker shades of
gray correspond to Co rich regions.\label{patterns}}
\end{figure}

We used a two-dimensional lattice with $450^2$ cells and periodic boundary conditions. The
order parameter represents the difference between the average concentration of Cu and
Co in the alloy, $\psi = \psi_{\mbox{\scriptsize{Cu}}} - \psi_{\mbox{\scriptsize{Co}}}$. Our initial condition is the
homogeneous system with $\psi(\vec{r})=\psi_0+\delta (\vec{r})$, where $\delta (\vec{r})$ is a small random 
fluctuation uniformly distributed in the interval $[-0.005,0.005]$. For a Cu$_{90}$Co$_{10}$ alloy,
 $\langle\psi_{\mbox{\scriptsize{Cu}}}\rangle=0.9$ and $\langle\psi_{\mbox{\scriptsize{Co}}}\rangle=0.1$ (average values of concentration),
so we have $\psi_0=0.8$.
For such values of concentration the system is in a metastable state after the
quench, so phase ordering proceeds via nucleation, and
segregation of phases can occur only if nucleation centers are present. In our
simulation nucleation centers consisting of four sites each with $\psi=-0.5$,
adding up to about $10\%$ of the lattice were randomly placed. For a
longer discussion about off-critical quench in CDS models see \cite{oopu2}. We used $D=0.5$
and our choice of map is
\begin{equation}
 f(x) = {\cal A} \tanh x  \,
\end{equation}
where ${\cal A}$ is a measure of the quench depth (${\cal A}>1$ corresponds to a
homogeneous mixture). In our simulations we use ${\cal A}=1.3$.
Figure~\ref{patterns} shows the snapshots of the lattice
after 100, 1000, 5000, 10000 iterations.
The color black is associated to the presence of Co, so, according to our definition of
$\psi$, the grains have $\psi<0$. The time evolution of the pattern is driven by
the gradient of the chemical potential, leading to domain coarsening and coalescence, so that
patterns obtained after different number of iterations have distinct distributions of grain sizes
and shapes, leading to distinct magnetic properties. As the
segregation process proceeds, grains coalesce and grow, such that the number of
grains decrease, as they become larger. In real samples such evolution of the
grain structure is possible only with a thermal treatment.
In this way, we are going to compare patterns simulated at later times with
experimental samples that have been annealed at higher temperatures.
\subsection{Analysis of the grain structure} \label{grains}
\begin{figure}
\includegraphics{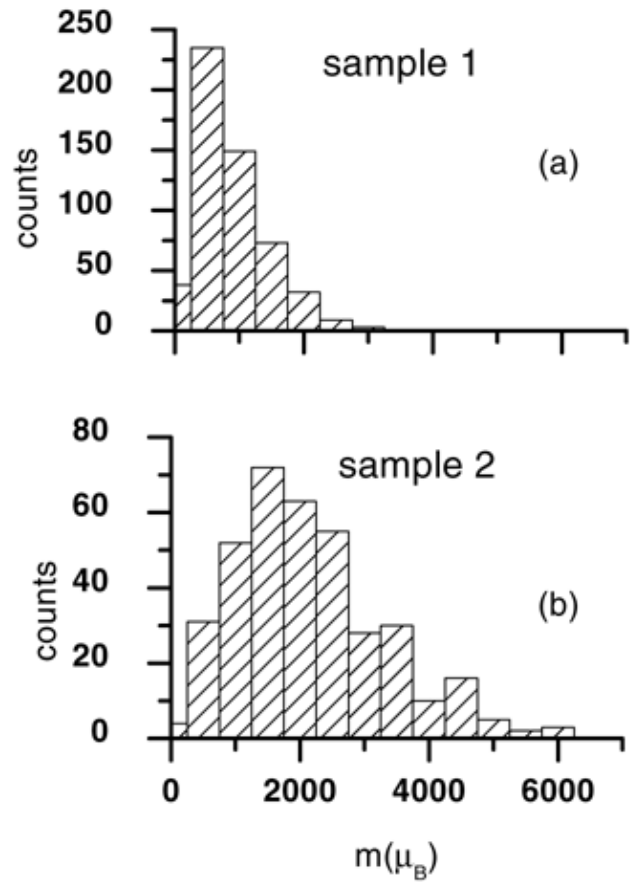}
\caption{Histograms for the patterns after 1000, (a), and 10000, (b), iterations shown in Fig. \ref{patterns}.
\label{histograms}
}

\end{figure}

We have used the standard algorithm developed by Hoshen and
Kopelman \cite{cluster} to obtain the list of cells belonging to each grain. In
order to check the form of the grain size distribution we have calculated the
histograms corresponding to granular patterns after given number of iterations of
Eq. \ref{cdsmodel} in terms of the magnetic moments $m$ of the grains. As explained
below, the magnetic moments are proportional to the number of sites in each cluster
with $\psi<0$.  Figure ~\ref{histograms} shows the histograms after 1000 (Sample 1), and
10000 (Sample 2) iterations, showing that as time goes on there is a shift to larger values of
magnetization, due to the coalescence of grains.
In order to compare the distribution obtained in simulated samples to the usually
adopted log-normal distribution, we have fitted the above histograms with the
curve
\begin{equation}
P(m)\mbox{d}m = \frac{A}{(\sqrt{2 \pi})w m}\exp\left[-\frac{\left(\ln\frac{m}{m_c}\right)^2}{2w^2}\right]\mbox{d}m \;,
\label{eqlognormal}
\end{equation} 
where $P(m)$ is the probability density to find the grain magnetic moment between $m$ and
$m+\mbox{d}m$, $m_c$ is the most probable value of $m$, and $w$ is the width of the distribution.

The resulting fitting parameters are summarized in Table \ref{lognormal}.
\begin{center}
\begin{table}
\begin{tabular}{|c|c|c|c|c|}  \hline
sample&	$m_c(\mu_B)$	&	$w$	&	$A(\mu_B)$
\\ \hline \hline
$1$	&  	$823$	&   	$0.80$ &$116\times 10^3$	\\ \hline
$2$	&	$2091$&	$0.63$ &$189\times 10^3$	\\ \hline
\end{tabular}
\caption{Parameters resulting from the fitting of a Log-normal distribution defined by Equation (\ref{eqlognormal}) to
the histograms of Fig.~ \ref{histograms}.\label{lognormal}
}
\end{table}
\end{center}

Cobalt grains have fcc symmetry \cite{fcc}, so it is reasonable to consider that
the dominant contribution to the anisotropy energy comes from the grains shape.
Observing the pattern formed after
1000 iterations (Fig. \ref{patterns}.(b)), for example, we notice essentially two kinds of
grain shapes, one almost
circular, and another elongated. Once the grains are labeled we can obtain information about their shape in order to
calculate the direction of the easy magnetization axis.  
The method for calculating this direction is quite simple, it consists in
calculating the position of the central site of the {\it i-}th grain, and then
summing up the position vectors of each site related to the position of this
central site. The resultant vector direction is the easy axis direction,
$\widehat{e}_{i}$.
\begin{figure}
\includegraphics{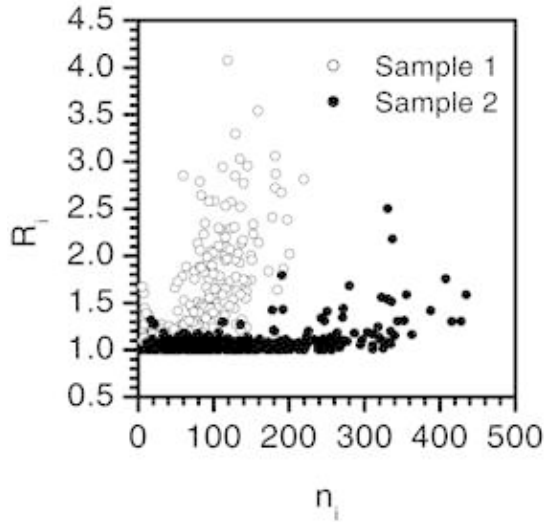}
\caption{Scatter plot of the anisotropy, measured as the ration $R_i$ between the
larger and the smaller dimension  of the $i$-th grain, as a function of the number
of sites, $n_i$ in the grain. It is clear from this plot that the segregation process 
 leads to larger and rounder grains, and that anisotropy is correlated
to grain size. We expect the same behavior in real systems, that is, annealing will
generate a system with larger but less anisotropic grains.
 \label{ratiosize}
}
\end{figure}

Finally we have translated our arbitrary units to physical ones, using a real Cub$_{90}$Co$_{10}$
sample for which the average values of the grain magnetic moment and anisotropy constants are known.
The determination of those experimental values is explained in Section \ref{exp}.
The first step was to associate a magnetization value to each grain.  Assuming that each grain
magnetization is proportional to its size, we can adjust our results by comparing them to
experimental data. Of course one serious difficulty is that the whole simulation is 2d,
so actually we have areas and not volumes for the grains.

 The average grain size  is just the average number of lattice sites
of the grains, $\overline{n}$, and was associated to
the average experimental magnetic moment $m_{exp}$ of a Cub$_{90}$ Co$_{10}$ sample as
\begin{equation}
\overline{n}f_{m} = m_{expo}\;\;,
\end{equation}
so that the magnetic moment of the $i$-th simulated grain  is then $m_i=n_if_{m}$, where $n_i$ is
the number of sites of that grain, and $f_{m}$ the conversion factor in units of $m_{exp}$.

We also need the conversion factor for the anisotropy constant $K$. We used the fact that, for cobalt,
when $R_i=3.5$ the shape anisotropy constant is equal to $K_{exp}=4.5\times 10^6\;$erg/cam$^3\;\;$ \cite{cullity}.
With this it is possible to define the conversion factor $f_K$ as
\begin{equation}
f_K=\frac{K_{expo}}{3.5}\label{fk}
\end{equation}
so that, for each grain, the anisotropy constant in the same units of $K_{exp}$ may
be calculated as $K_i=f_K R_i$.

The properties of the samples used in the Monte Carlo simulations are summarized
in Table \ref{samples}.
\begin{center}
\begin{table}
\begin{tabular}{|c|c|c|c|c|} \hline
sample&number of iterations& $N_g$ & $\overline{m}(\mu_B)$& $\overline{K}(10^6\;$erg/cam$^3)$\\ \hline\hline 1&1000&539&869&1.58\\\hline
2&10000&371&2140&1.58\\\hline
\end{tabular}
\caption{Properties of the simulated Cub$_{90}$Co$_{10}$ samples used in the Monte Carlo simulations.
$N_g$ is the total number of grains in a sample, $\overline{m}(\mu_B)$ is the average magnetic moment of the
grains and $\overline{K}$ is the average anisotropy constant of the sample.
\label{samples}
}
\end{table}
\end{center}

\subsection{Experiment}\label{exp}
Continuous ribbons of Co$_{10}$Cub$_{90}$ were obtained by melt spinning in AR atmosphere using a Cub-AR drum. The as
cast
material was subjected to furnace annealing at temperature range 400-600$\;^o$C for 60 min, generating samples with different
 nanostructures. Details of the sample structure and  how it is affected by the thermal treatment can be seen in
 reference \cite{yu}.  The magnetization measurements were performed on a commercial Quantum Design PPMS system with temperatures in the range
2-300 K, for different values of the applied field $H$. In each curve the sample was cooled to 2 K in zero field. After applying a
magnetic field, the magnetization was measured with increasing temperature up to 300 K (ZFC magnetization). Subsequently, the
magnetization was measured for decreasing temperature in the same field down to 2 K (FC magnetization). The magnetic field was applied
by a superconducting magnet operating in persistent mode and the total time of each complete ZFC and FC curves was of 7 hours with a
rate of temperature sweep of 1.5 K/min. 

The moment distribution of the samples was obtained by fitting a sum of Langevin functions weighted by a log-normal moment
distribution in the $m(H)$ curve measured at $T = 300\;$K$\;$ \cite{ferrari}. The saturation magnetization was obtained by extrapolation of
the curve $m(1/H)$ to zero temperature.
%
\section{Monte Carlo simulation}\label{sec:Md}
A ferromagnetic particle becomes a monodomain when its linear size is below a critical value $D_c$
determined by the minimization of the total energy, including magnetostatic,
exchange and anisotropy contributions \cite{bertotti}. Below this critical size, the energy associated to the creation
of magnetic domain walls is larger than the decrease in the magnetostatic energy due to the smaller total magnetization.
 Such monodomain
ferromagnetic particles can be viewed as large magnetic units, each having a magnetic moment
of thousands of Bohr magnetons.  Usually, in low concentration alloys, neighboring
particles are separated by $10 - 30$ nm, and direct exchange, as well as indirect
(like RKKY) between particles is neglected \cite{altbir1}.
Thus, the magnetic properties of an assembly of nanoparticles are determined by the dipolar
interaction energy between the particles along with thermal and magnetic anisotropy energies.
In this paper we consider only the latter effect.
Experiments conducted on magnetic nanoparticles show irreversible magnetic behavior below a
irreversibility line $T_{irr}(H)$.  In particular, the ZFC and FC magnetization curves do not
coincide, and magnetic hysteresis appears. In such systems the origin
of irreversibility is the interplay between thermal energy and some energy barrier, which hinders
relaxation towards equilibrium. The magnetic irreversibility in nanoparticles is
conventionally associated with the energy required for a particle moment reorientation,
overcoming a barrier due to shape, magnetoelasticity, or crystalline anisotropy \cite{bertotti}.

 Our system consists of $N_g$ magnetic monodomain
particles, whose sizes, shapes and anisotropy constants are obtained by means of the previously
described procedure. Each particle is described by its magnetic moment
$\overrightarrow{m}_{i}$, the direction of the easy  magnetization axis,
$\widehat{e}_{i}$ and its anisotropy constant $K_i$. All those quantities result from the CDS simulation of 
the granular structure. In the presence of an external
magnetic field $\overrightarrow{H}$, the Hamiltonian of our system can be written
as
\begin{equation}
\mathcal{H}=\sum_{i}\left[-\overrightarrow{m}_{i}\cdot \overrightarrow{H}-
\kappa_{i} \left( \frac{\overrightarrow{m}_{i}\cdot \widehat{e}_{i}}{m_{i}}\right)^{2} \right] \; \;,
\label{hamiltonian}
\end{equation}
where $\kappa_{i} = K_{i}N_{i}$, $N_{i} $ being the number of atoms of grain {\it i}.

The ZFC-FC curves correspond to nonequilibrium states of the system, therefore they are dependent
on the temperature variation rate. In terms of a Monte Carlo simulation this means that we
cannot wait too long at each value of temperature, and the usual mechanism of time averaging
instead of ensemble averaging is not valid. The estimation of the number of Monte Carlo steps
in each temperature was completely arbitrary and it has no obvious relation to the actual time used in
experiments. However, as we keep this number constant in our simulations we can at least
guarantee that the cooling and heating rates are equal in all simulations and our conclusions will
be valid at these rates. The
averaging was done over 200 independent but statistically equivalent sweeps.
For the ZFC curve we started from a configuration where the magnetic moments of the grains
were randomly chosen and $H = 0$. The simulation proceeded by turning on the external field
(typically 10-100 Oe) in the $x-$direction at low temperature ($T=2\;$K). The $N_g$ grains were then
sequentially chosen, and had their magnetic moment rotated by an angle sorted from a uniform distribution. The change in
energy ($\Delta E$) was calculated and the rotation accepted with probability
$p=\mbox{min}[1,\exp(-\Delta E/k_B T)]$. The update of the $N_g$ particles was repeated  100 times for the
initial value of temperature, and then the temperature was increased by 2 K, and so on.
At $T = 200\;$K we kept the last configuration of magnetic moments and used it as the initial
configuration for the FC curve, following the same procedure used in the ZFC curve, only
decreasing the temperature 2 K  each step.

%
\section{Results}\label{sec:Res}
The analysis of ZFC curves usually involves two temperatures $T_M$ and $T_{irr}$
defined as the temperature at the maximum and the temperature above which the system shows thermodynamic
equilibrium properties corresponding to a superparamagnetic behavior, respectively \cite{dormann}. For
$T<T_{irr}$ irreversibility impedes the coincidence of the ZFC and FC curves.
For $T>T_{irr}$, the relaxation time for the magnetization of the largest particle is
much smaller than the typical measuring time, and the ZFC and FC curves coincide. Our goal
is to understand how these temperatures are affected by the distribution of
sizes and shapes of the particles, magnitude of the applied
field and annealing of the grains.
\subsection{Influence of the distribution of grain sizes and shapes}
In order to compare the effect of the distribution of grain sizes and anisotropies on the ZFC-FC
curves, we performed a series of simulations using different types of distributions namely, uniform,
Gaussian and realistic (obtained from CDS simulation) for the sizes, anisotropy axis sorted from a uniform
distribution of directions, and obtained from the CDS simulation,
anisotropy constants equal to the average value, and calculated from the simulated alloy. One important point in the choice
of the anisotropy constants $K_i$, is that they are intensive quantities.
In order to keep the total anisotropy energy independent of the choice of the $K$ distribution,
we calculated the average value of the anisotropy constant using the grain sizes as weights.
 The results below show the combination of
those possibilities for the same sample under an applied external field  $H = 0.1\;$KOe. 
All curves show the behavior of the reduced average magnetization per grain, $m/m_s$, where
$m_s$ is the saturation value of the sample.

We begin our study by considering a sample with the same number of grains, average magnetic
moment and average anisotropy constant of sample 1, as defined in Table \ref{samples}, but with
arbitrarily chosen size distributions.

\begin{center}
\begin{figure}
\includegraphics{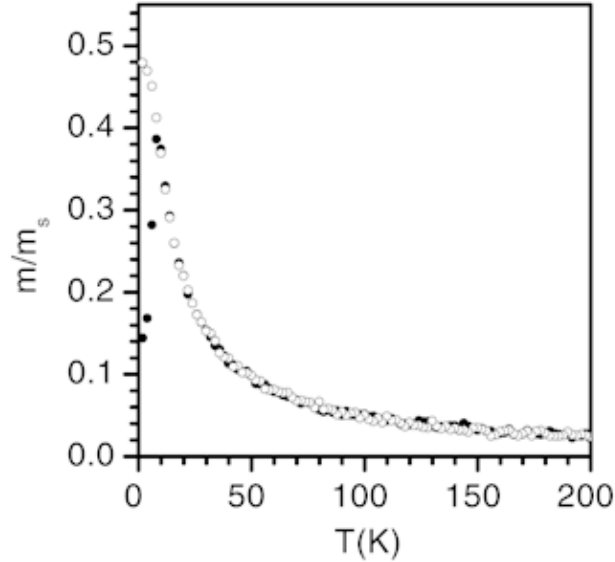}
\caption{Simulated ZFC-FC curves for a sample of 539 grains with constant
magnetic moment ($=869 \mu_B$) and anisotropy constant ($K=1.58\;$erg/cm$^3$) and anisotropy axis sorted from a uniform distribution. Since all particles have the same size, $T_{irr}=T_M(=8$K in this case), as expected.\label{dist1} }
\end{figure}
\end{center}

First we assign a constant value of magnetization, $m_i=\overline{m} = 869 \mu_B$, and uniform
anisotropy constant, $K_i=\overline{K} = 1.58\times 10^6\;$erg/cm$^3$, to all grains 
and  anisotropy axis sorted from a uniform distribution. These choices represent a system in which
all grains have the same size and shape, with anisotropy axis determined by the random alignment 
of the crystalline axis and the magnetic field. Figure ~\ref{dist1}  shows the results for this system. We
notice that $T_{irr} = T_M = 8\;$K and that the peak of the ZFC is sharp.

\begin{figure}
\includegraphics{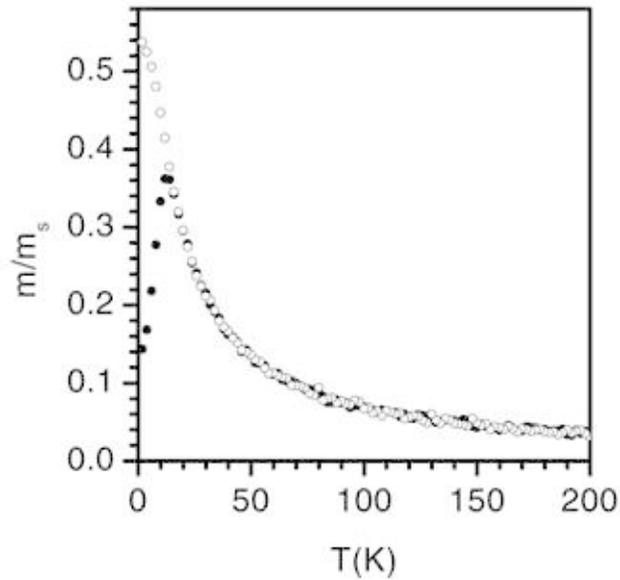}
\caption{Simulated ZFC-FC curves for a sample of 539 grains with Gaussian distribution
of magnetic moments ($\overline{m}=869\;\mu_B$ and $\sigma=550\;\mu_B$). All grains
have the same anisotropy constant ($K=1.58\;$erg/cm$^3$) and the anisotropy axis are
sorted from a uniform distribution. Although grains of different sizes are present, we still have $T_{irr}=T_M$,
indicating that the existence of a size distribution may not be the sole reason for having
$T_{irr}>T_M$. Here $T_{irr}=T_M=12\;$K.\label{dist2}
}
\end{figure}

Next we improve the description by considering a system with magnetizations sorted from a Gaussian
distribution. In order to have a distribution quantitatively similar to the simulated one, we
have chosen it to be centered at $\overline{m} = 869 \mu_B$ with variance corresponding
to the width of the distribution of sample 1, that is, $\sigma = 550 \mu_B$. We still keep the
constant value of anisotropy $K$, i.e. all particles with the same shape, and 
anisotropy axis sorted from the uniform distribution. With these choices we have a system 
with particles of different sizes, but uniform shape, and with easy magnetization axis
determined by the grain crystalline axis, as before. 
As can be seen
in Figure ~\ref{dist2}, the existence of particles with different sizes is not
a sufficient condition for having $T_{irr}\neq T_M$ as observed in experimental curves. A
symmetrical distribution of sizes leads to curves similar to the ones obtained from
a system of particles with uniform size in the sense that $T_{irr}=T_M$ in both cases.
However, because of the inclusion of a size distribution, we have now 
grains larger than in the previous case, and this is reflected in the larger values of those
 temperatures. Bigger particles with higher
magnetizations weight more in the ZFC curve than the smaller ones, and because of its bigger
size they yield  bigger values of $T_{irr}$. Therefore, the position of
$T_{irr}$ and $T_M$ shift to higher
temperatures and  now $T_{irr}=T_M=12\;$K.

\begin{figure}
\includegraphics{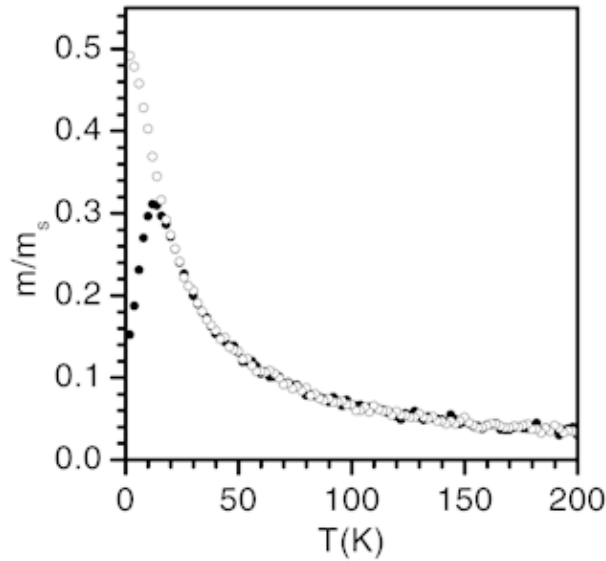}
\caption{Simulated ZFC-FC curves for a sample of 539 grains with a realistic distribution
of magnetic moments from the CDS simulation
in which  $\overline{m}=869\;\mu_B$. As in Fig. \ref{dist2} all grains
have the same anisotropy constant ($=1.58\;$erg/cm$^3$) and the anisotropy axis are sorted from a uniform distribution.
In this case $T_{irr}=18\;$K and $T_M=13\;$K.\label{dist3}
}
\end{figure}

The effect of the distribution profile can be clearly seen if we use the
asymmetrical distribution generated by the CDS simulation, together with constant
$K$ and uniformly distributed anisotropy axis.
 Figure ~\ref{dist3} shows the resulting curves for this case, where $T_{irr}>T_M$
($T_{irr}=18\;$K and $T_M=13\;$K)  and both are larger than the values obtained in the
previous case, with the symmetrical distribution. Moreover the maximum value of
$m/m_s$ in the ZFC curve is lower as compared to the symmetrical distribution
curve.  In this case, the number of smaller particles is bigger, as compared to
the monodisperse case. Some small particles 
are becoming disordered at the
temperature of the maximum magnetization of the sample, therefore, the
magnetization peak in the ZFC curve is lower. Also for this reason,
$T_{irr}$ is larger.

Finally we use the complete information from the CDS simulation, that is,
distribution of sizes, axis and anisotropy constants calculated from the simulated
grains. In this case, $T_{irr}=36\;$K and $T_M=8\;$K.

\begin{figure}
\includegraphics{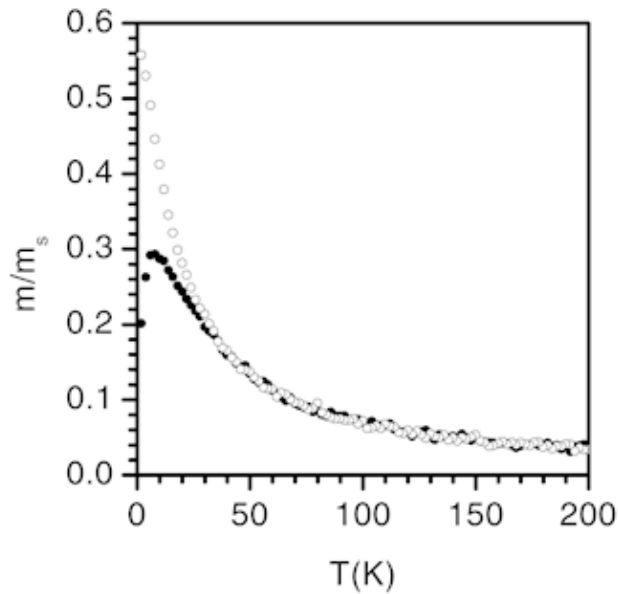}
\caption{Simulated ZFC-FC curves for sample 1 of Table \ref{samples}. The
introduction of a realistic distribution of anisotropy constants generates a ZFC
curve more similar to an experimental one (see Fig.~\ref{anneal}),  as
the ZFC and FC curves separate at a temperature well below $T_M$.  Here $T_{irr}=36\;$K $T_M=8\;$K.\label{dist5} }
\end{figure}

From now on, all the simulated curves use the structural properties of the simulated alloy, as explained in Section \ref{grains}.
\subsection{Effect of annealing}

Even for non interacting grains, annealing is capable of drastically changing the
ZFC-FC curves.  As the sample is brought to a temperature closer to the phase separation
critical temperature, diffusion is enhanced and fabrication defects are relaxed. The overall
effect is an isotropic sample with larger and fewer grains. For this reason we have chosen an annealed sample as our starting reference system. As the thermal treatment is repeated,
grains grow larger and coalescence further decreases the number of grains. It is possible to achieve the
same result in the simulated alloy simply by iterating more times the time evolution
rule defined by equation (\ref{cdsmodel}). For this reason we assume that sample 2, collected 9000 iterations after sample
1 was collected, represents an annealed sample, as compared to sample 1. Using the same conversion factors defined for sample 1 in s
ample 2, we guarantee consistency in the values of $m_i$ and $K_i$.

\begin{figure}
\includegraphics{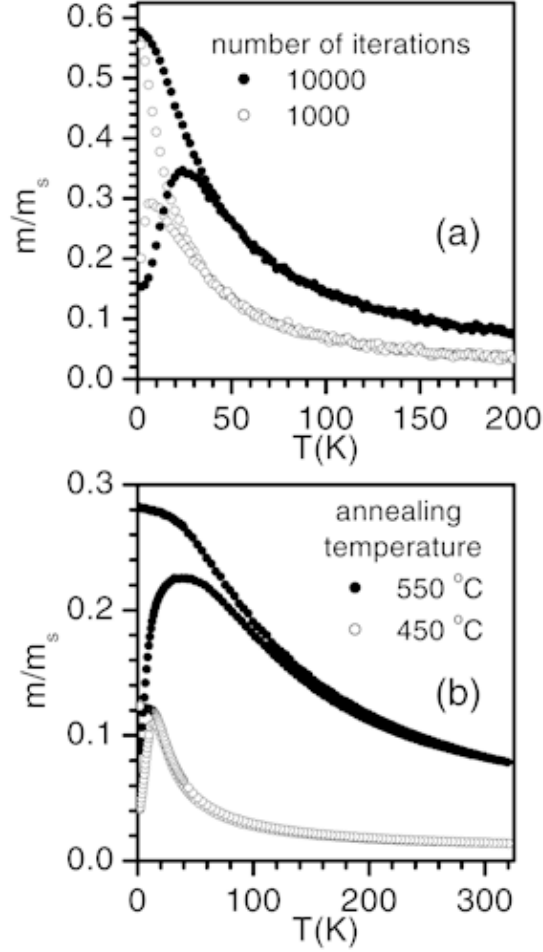} \caption{(a) Simulated ZFC-FC
curves for samples 1 and 2. (b) Experimental curves for a sample annealed for an
hour at 450$\;^o$C and 550$\;^o$C. Higher annealing temperatures facilitate the
process of segregation, so we have a correspondence to a simulated sample with a
larger number of iterations of the time evolution equation of the segregation process, Eq. (\ref{cdsmodel}).\label{anneal} }
\end{figure}

As expected
for a sample with larger grains, both $T_M$ and $T_{irr}$ are larger, and the maximum
of both curves are higher.  While a direct comparison is difficult, we can see
that there is a qualitative agreement with experimental results, as illustrated in
Fig. \ref{anneal}.(b). Because of the existence of bigger grains in sample 2, the
mean value of the magnetization increases; therefore $T_{irr}$
increases and also the low temperature limit of $m/m_s$ on the FC curve.\\
\subsection{Effect of external field}
Here we consider the effect of
varying the magnitude of the applied magnetic field on sample 1. Figure~\ref{fields}
illustrates our results for fields in the range 50-1000 Oe.  Simulated curves have the
same qualitative behavior found in experiments, as depicted in Figure~\ref{fieldsexp}. That
is, both ZFC and FC curves are dislocated to larger values of $m/m_s$ as the field increases, and
$T_M$ and $T_{irr}$ decrease. Table \ref{tabledif} shows an
estimation of the values of $T_{irr}$ and $T_M$ for the simulated curves.
\begin{center}
\begin{figure}
\includegraphics{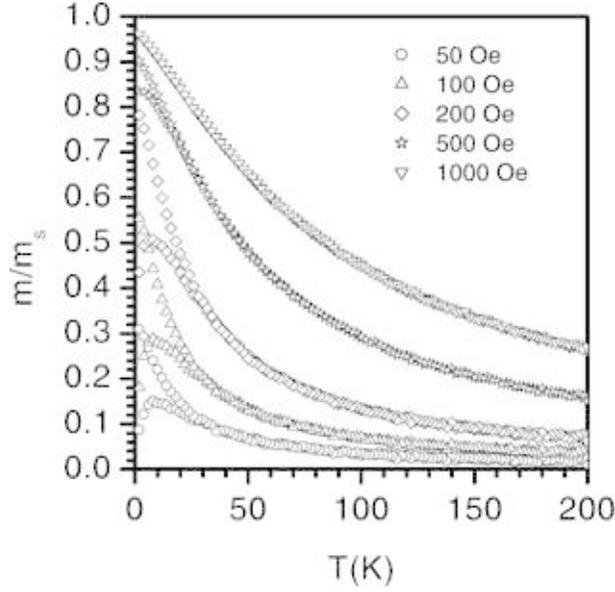}
\caption{Simulated ZFC-FC curves for sample 1 with different values of the applied field.
Both $T_{irr}$ and $T_M$ decrease with increasing field as can be seen in Table \ref{tabledif}.\label{fields}}
\end{figure}
\begin{center}
\begin{figure}
\includegraphics{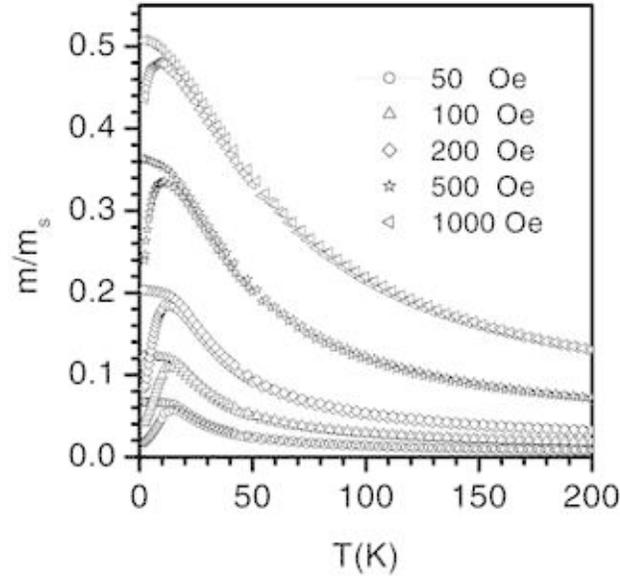}
\caption{Experimental ZFC-FC curves for a sample annealed at 450$\;^o$C.
Corresponding values of $T_{irr}$ and $T_M$ are listed in Table~\ref{fields}.\label{fieldsexp}}
\end{figure}
\end{center}
\begin{table}
\begin{tabular}{|c|c|c|c|c|} \hline
&\multicolumn{2}{c|}{Simulation}&\multicolumn{2}{c|}{Experiment}\\ \hline
field (Oe)&$T_{irr}(K)$&$T_M$(K)&$T_{irr}$(K)&$T_M$\\ \hline\hline
50&36&8&39&16\\ \hline
100&36&8&38&15\\\hline
200&32&8&35&14\\\hline
500&26&4&22&11\\\hline
1000&14&2&19&9\\\hline \end{tabular}
\caption{Values of $T_{irr}$ and $T_M$ for the curves in Fig.~\ref{fields} and \ref{fieldsexp}.\label{tabledif} }
\end{table}
\end{center}

Both values of temperature reflect the competition between thermal excitation and the energy barrier
between the two axial orientations.
$T_{irr}$ is strongly influenced by the external field. As $T_{irr}$ is related with equilibrium, i.e,
it is the temperature above which the system behaves like
a superparamagnet, by increasing the external field, the relaxation time diminishes.
For superparamagnetic
 systems,   the relaxation time at temperature $T$ is given by
$\tau = \tau_0 \exp\left(\frac{\Delta E}{k_BT}\right)$, where  $\tau_0$ is a constant of the order of $10^{-9}\;$ s and $\Delta E$
 is the energy barrier that each particle has to overcome to minimize its
 energy. Because the
 energy barrier  decreases under  an external magnetic field,   so do $T_M$ and $T_{irr}$, and
the equilibrium state can be reached at lower temperatures.

\section{Conclusions}
There are several difficulties involved in the direct comparison of
simulated and experimental curves, mostly because we are dealing with
nonequilibrium states. The first issue is how to identify a simulated Co$_{x}$Cu$_{1-x}$
sample with a real one. As explained above, the CDS simulation deals with
densities, and the space-time scale is undefined. As a zero order
approximation we can force the simulated alloy to have the same
average value of magnetic moment, but the variance of the grain
size distribution in the simulation cannot be adjusted to be exactly the same
as in the real system.
Besides all this, there is also the problem of
relating Monte Carlo steps with cooling and heating rates of the ZFC and FC curves,
as discussed above.
In summary, even though we have used experimental parameters to
keep our simulated curves as close as possible to the real ones, a direct
comparison of them is too ambitious. However, qualitative comparisons to determine the
effect of each of the parameters of our model are valid, and the importance of
relating anisotropy, size and shape distributions is stated.

The existence of a temperature $T_{irr}\neq T_M$ marking the irreversible portion of the ZFC curve
is already well understood to be a consequence of a grain size distribution \cite{dormann}. What we
observed in this work is that the sole existence of size distribution is not sufficient to
produce a simulated ZFC curve similar to the experimental one.  For symmetrical size
distribution of the magnetic particles, $T_M$ coincide with $T_{irr}$, within numerical accuracy as can be seen in Fig.~\ref{dist2}.
However, for non symmetrical particle size distributions, $T_{irr}> T_M$ as shown in Fig.~\ref{dist3}.
The introduction of a distribution of anisotropy constants is a delicate point. As can be seen in
Fig. \ref{ratiosize}, shape and size are correlated, so one cannot independently sort the values 
of magnetic moments and anisotropy constants, even when using the correct distributions.
This effect has been so far neglected in simulations
of granular systems.
The qualitative comparison of the ZFC-FC curves corresponding to samples that have
evolved in time with experimental curves for thermally treated samples reinforces this
idea. As the binary alloy relaxes towards its equilibrium configuration, fewer and larger grains are present,
and only the larger ones show appreciable anisotropy as shown in the scatter
plot of anisotropy
as a function of particle size for samples obtained after times (Fig.~\ref{ratiosize}) for samples 1 and
2 obtained after different number of iterations.

\begin{acknowledgments}
A bilateral project Vitae/Fundaci\'{o}n Andes is
acknowledged by the authors. In Chile the groups received
financial support from FONDECYT under grants \# 1010127 and
1990812, and Millennium Science Nucleus ``Condensed Matter
Physics'' P99-135F. In Brazil, the authors acknowledge the support
from FAPESP, FAPERJ, CAPES and CNPq.

\end{acknowledgments}


\begin{thebibliography}{30}
\expandafter\ifx\csname natexlab\endcsname\relax\def\natexlab#1{#1}\fi
\expandafter\ifx\csname bibnamefont\endcsname\relax
  \def\bibnamefont#1{#1}\fi
\expandafter\ifx\csname bibfnamefont\endcsname\relax
  \def\bibfnamefont#1{#1}\fi
\expandafter\ifx\csname citenamefont\endcsname\relax
  \def\citenamefont#1{#1}\fi
\expandafter\ifx\csname url\endcsname\relax
  \def\url#1{\texttt{#1}}\fi
\expandafter\ifx\csname urlprefix\endcsname\relax\def\urlprefix{URL }\fi
\providecommand{\bibinfo}[2]{#2}
\providecommand{\eprint}[2][]{\url{#2}}

\bibitem[{\citenamefont{Wei et~al.}(1994)\citenamefont{Wei, Krauss, and
  Fischer}}]{recording}
\bibinfo{author}{\bibfnamefont{S.~Y. C. M.~S.} \bibnamefont{Wei}},
  \bibinfo{author}{\bibfnamefont{P.~R.} \bibnamefont{Krauss}},
  \bibnamefont{and} \bibinfo{author}{\bibfnamefont{P.~B.}
  \bibnamefont{Fischer}}, \bibinfo{journal}{J. Appl. Phys.}
  \textbf{\bibinfo{volume}{76}}, \bibinfo{pages}{6673} (\bibinfo{year}{1994}).

\bibitem[{\citenamefont{El-Hilo et~al.}(1994)\citenamefont{El-Hilo, O'Grady,
  and Chantrell}}]{el-hilo}
\bibinfo{author}{\bibfnamefont{M.}~\bibnamefont{El-Hilo}},
  \bibinfo{author}{\bibfnamefont{K.}~\bibnamefont{O'Grady}}, \bibnamefont{and}
  \bibinfo{author}{\bibfnamefont{R.~W.} \bibnamefont{Chantrell}},
  \bibinfo{journal}{J. Appl. Phys.} \textbf{\bibinfo{volume}{76}},
  \bibinfo{pages}{6811} (\bibinfo{year}{1994}).

\bibitem[{\citenamefont{Gittleman et~al.}(1972)\citenamefont{Gittleman,
  Goldstein, and Bozowski}}]{gittleman}
\bibinfo{author}{\bibfnamefont{J.~I.} \bibnamefont{Gittleman}},
  \bibinfo{author}{\bibfnamefont{Y.}~\bibnamefont{Goldstein}},
  \bibnamefont{and} \bibinfo{author}{\bibfnamefont{S.}~\bibnamefont{Bozowski}},
  \bibinfo{journal}{Phys. Rev. B} \textbf{\bibinfo{volume}{5}},
  \bibinfo{pages}{3609} (\bibinfo{year}{1972}).

\bibitem[{\citenamefont{Ferrari et~al.}(1997)\citenamefont{Ferrari, da~Silva,
  and Knobel}}]{ferrari}
\bibinfo{author}{\bibfnamefont{E.~F.} \bibnamefont{Ferrari}},
  \bibinfo{author}{\bibfnamefont{F.~C.~S.} \bibnamefont{da~Silva}},
  \bibnamefont{and} \bibinfo{author}{\bibfnamefont{M.}~\bibnamefont{Knobel}},
  \bibinfo{journal}{Phys. Rev. B} \textbf{\bibinfo{volume}{56}},
  \bibinfo{pages}{6086} (\bibinfo{year}{1997}).

\bibitem[{\citenamefont{Hickey et~al.}(1995)\citenamefont{Hickey, Howson, Musa,
  and Wiser}}]{hickey}
\bibinfo{author}{\bibfnamefont{B.~J.} \bibnamefont{Hickey}},
  \bibinfo{author}{\bibfnamefont{M.~A.} \bibnamefont{Howson}},
  \bibinfo{author}{\bibfnamefont{S.~O.} \bibnamefont{Musa}}, \bibnamefont{and}
  \bibinfo{author}{\bibfnamefont{N.}~\bibnamefont{Wiser}},
  \bibinfo{journal}{Phys. Rev. B} \textbf{\bibinfo{volume}{51}},
  \bibinfo{pages}{667} (\bibinfo{year}{1995}).

\bibitem[{\citenamefont{Wiser}(1996)}]{wiser}
\bibinfo{author}{\bibfnamefont{N.}~\bibnamefont{Wiser}}, \bibinfo{journal}{J.
  Magn. Magn. Mater.} \textbf{\bibinfo{volume}{159}}, \bibinfo{pages}{119}
  (\bibinfo{year}{1996}).

\bibitem[{\citenamefont{Dimitrov and Wysin}(1996)}]{dimitrov}
\bibinfo{author}{\bibfnamefont{D.~A.} \bibnamefont{Dimitrov}} \bibnamefont{and}
  \bibinfo{author}{\bibfnamefont{G.~M.} \bibnamefont{Wysin}},
  \bibinfo{journal}{Phys. Rev. B} \textbf{\bibinfo{volume}{54}},
  \bibinfo{pages}{9237} (\bibinfo{year}{1996}).

\bibitem[{\citenamefont{Allia et~al.}(1995)\citenamefont{Allia, Knobel,
  Tiberto, and Vinai}}]{allia}
\bibinfo{author}{\bibfnamefont{P.}~\bibnamefont{Allia}},
  \bibinfo{author}{\bibfnamefont{M.}~\bibnamefont{Knobel}},
  \bibinfo{author}{\bibfnamefont{P.}~\bibnamefont{Tiberto}}, \bibnamefont{and}
  \bibinfo{author}{\bibfnamefont{F.}~\bibnamefont{Vinai}},
  \bibinfo{journal}{Phys. Rev. B} \textbf{\bibinfo{volume}{52}},
  \bibinfo{pages}{15398} (\bibinfo{year}{1995}).

\bibitem[{\citenamefont{Chantrell et~al.}(2000)\citenamefont{Chantrell,
  Walmsley, Gore, and Maylin}}]{Chantrell}
\bibinfo{author}{\bibfnamefont{R.~W.} \bibnamefont{Chantrell}},
  \bibinfo{author}{\bibfnamefont{N.}~\bibnamefont{Walmsley}},
  \bibinfo{author}{\bibfnamefont{J.}~\bibnamefont{Gore}}, \bibnamefont{and}
  \bibinfo{author}{\bibfnamefont{M.}~\bibnamefont{Maylin}},
  \bibinfo{journal}{Phys. Rev. B} \textbf{\bibinfo{volume}{63}},
  \bibinfo{pages}{024410} (\bibinfo{year}{2000}).

\bibitem[{\citenamefont{Pike et~al.}(2000)\citenamefont{Pike, Roberts, and
  Verosub}}]{Pike}
\bibinfo{author}{\bibfnamefont{C.~R.} \bibnamefont{Pike}},
  \bibinfo{author}{\bibfnamefont{A.~P.} \bibnamefont{Roberts}},
  \bibnamefont{and} \bibinfo{author}{\bibfnamefont{K.~L.}
  \bibnamefont{Verosub}}, \bibinfo{journal}{J. Appl. Phys.}
  \textbf{\bibinfo{volume}{88}}, \bibinfo{pages}{967} (\bibinfo{year}{2000}).

\bibitem[{\citenamefont{Gunton et~al.}(1983)\citenamefont{Gunton, Miguel, and
  Sahini}}]{dg8}
\bibinfo{author}{\bibfnamefont{J.~D.} \bibnamefont{Gunton}},
  \bibinfo{author}{\bibfnamefont{M.~S.} \bibnamefont{Miguel}},
  \bibnamefont{and} \bibinfo{author}{\bibfnamefont{P.~S.} \bibnamefont{Sahini}}
  (\bibinfo{publisher}{Academic Press, New York}, \bibinfo{year}{1983}),
  vol.~\bibinfo{volume}{8}, chap.~\bibinfo{chapter}{2}, p.
  \bibinfo{pages}{269}.

\bibitem[{\citenamefont{Allen and Cahn}(1979)}]{allencahn1}
\bibinfo{author}{\bibfnamefont{S.~M.} \bibnamefont{Allen}} \bibnamefont{and}
  \bibinfo{author}{\bibfnamefont{J.~W.} \bibnamefont{Cahn}},
  \bibinfo{journal}{Acta Metall.} \textbf{\bibinfo{volume}{27}},
  \bibinfo{pages}{1085} (\bibinfo{year}{1979}).

\bibitem[{\citenamefont{Binder}(1974)}]{binder1}
\bibinfo{author}{\bibfnamefont{K.}~\bibnamefont{Binder}}, \bibinfo{journal}{Z.
  Pysik.} \textbf{\bibinfo{volume}{267}}, \bibinfo{pages}{313}
  (\bibinfo{year}{1974}).

\bibitem[{\citenamefont{{d}a Silva et~al.}(1999)\citenamefont{{d}a Silva,
  Ferrari, and Knobel}}]{knobel1}
\bibinfo{author}{\bibfnamefont{F.~C.~S.} \bibnamefont{{d}a Silva}},
  \bibinfo{author}{\bibfnamefont{E.~F.} \bibnamefont{Ferrari}},
  \bibnamefont{and} \bibinfo{author}{\bibfnamefont{M.}~\bibnamefont{Knobel}},
  \bibinfo{journal}{J. Appl. Phys.} \textbf{\bibinfo{volume}{86}},
  \bibinfo{pages}{7170} (\bibinfo{year}{1999}).

\bibitem[{\citenamefont{Oono and Puri}(1987)}]{oopu1}
\bibinfo{author}{\bibfnamefont{Y.}~\bibnamefont{Oono}} \bibnamefont{and}
  \bibinfo{author}{\bibfnamefont{S.}~\bibnamefont{Puri}},
  \bibinfo{journal}{Phys. Rev. Lett.} \textbf{\bibinfo{volume}{58}},
  \bibinfo{pages}{836} (\bibinfo{year}{1987}).

\bibitem[{\citenamefont{Oono and Puri}(1988)}]{oopu2}
\bibinfo{author}{\bibfnamefont{Y.}~\bibnamefont{Oono}} \bibnamefont{and}
  \bibinfo{author}{\bibfnamefont{S.}~\bibnamefont{Puri}},
  \bibinfo{journal}{Phys. Rev. A} \textbf{\bibinfo{volume}{38}},
  \bibinfo{pages}{434} (\bibinfo{year}{1988}).

\bibitem[{\citenamefont{Oono and Yeung}(1987)}]{ooyeung1}
\bibinfo{author}{\bibfnamefont{Y.}~\bibnamefont{Oono}} \bibnamefont{and}
  \bibinfo{author}{\bibfnamefont{C.}~\bibnamefont{Yeung}}, \bibinfo{journal}{J.
  Stat. Phys.} \textbf{\bibinfo{volume}{48}}, \bibinfo{pages}{593}
  (\bibinfo{year}{1987}).

\bibitem[{\citenamefont{Shinozaki and Oono}(1991)}]{shinooo2}
\bibinfo{author}{\bibfnamefont{A.}~\bibnamefont{Shinozaki}} \bibnamefont{and}
  \bibinfo{author}{\bibfnamefont{Y.}~\bibnamefont{Oono}},
  \bibinfo{journal}{Phys. Rev. Lett.} \textbf{\bibinfo{volume}{66}},
  \bibinfo{pages}{173} (\bibinfo{year}{1991}).

\bibitem[{\citenamefont{Oono and Bahiana}(1988)}]{ooba1}
\bibinfo{author}{\bibfnamefont{Y.}~\bibnamefont{Oono}} \bibnamefont{and}
  \bibinfo{author}{\bibfnamefont{M.}~\bibnamefont{Bahiana}},
  \bibinfo{journal}{Phys. Rev. Lett.} \textbf{\bibinfo{volume}{61}},
  \bibinfo{pages}{1109} (\bibinfo{year}{1988}).

\bibitem[{\citenamefont{Mondello and Goldenfeld}(1990)}]{mogo1}
\bibinfo{author}{\bibfnamefont{M.}~\bibnamefont{Mondello}} \bibnamefont{and}
  \bibinfo{author}{\bibfnamefont{N.}~\bibnamefont{Goldenfeld}},
  \bibinfo{journal}{Phys. Rev. A} \textbf{\bibinfo{volume}{42}},
  \bibinfo{pages}{5865} (\bibinfo{year}{1990}).

\bibitem[{\citenamefont{Massunaga et~al.}(1997)\citenamefont{Massunaga,
  Paniconi, and Oono}}]{masscross}
\bibinfo{author}{\bibfnamefont{M.~S.~O.} \bibnamefont{Massunaga}},
  \bibinfo{author}{\bibfnamefont{M.}~\bibnamefont{Paniconi}}, \bibnamefont{and}
  \bibinfo{author}{\bibfnamefont{Y.}~\bibnamefont{Oono}},
  \bibinfo{journal}{Phys. Rev. E} \textbf{\bibinfo{volume}{56}},
  \bibinfo{pages}{723} (\bibinfo{year}{1997}).

\bibitem[{\citenamefont{Martins et~al.}(2000)\citenamefont{Martins, Morgado,
  Massunaga, and Bahiana}}]{swmm1}
\bibinfo{author}{\bibfnamefont{S.}~\bibnamefont{Martins}},
  \bibinfo{author}{\bibfnamefont{W.}~\bibnamefont{Morgado}},
  \bibinfo{author}{\bibfnamefont{M.}~\bibnamefont{Massunaga}},
  \bibnamefont{and} \bibinfo{author}{\bibfnamefont{M.}~\bibnamefont{Bahiana}},
  \bibinfo{journal}{Phys. Rev. E} \textbf{\bibinfo{volume}{61}},
  \bibinfo{pages}{4118} (\bibinfo{year}{2000}).

\bibitem[{\citenamefont{Puri and Oono}(1988)}]{puoo1}
\bibinfo{author}{\bibfnamefont{S.}~\bibnamefont{Puri}} \bibnamefont{and}
  \bibinfo{author}{\bibfnamefont{Y.}~\bibnamefont{Oono}},
  \bibinfo{journal}{Phys. Rev. A} \textbf{\bibinfo{volume}{38}},
  \bibinfo{pages}{1542} (\bibinfo{year}{1988}).

\bibitem[{\citenamefont{Hoshen and Kopelman}(1976)}]{cluster}
\bibinfo{author}{\bibfnamefont{J.}~\bibnamefont{Hoshen}} \bibnamefont{and}
  \bibinfo{author}{\bibfnamefont{R.}~\bibnamefont{Kopelman}},
  \bibinfo{journal}{Phys. Rev. B} \textbf{\bibinfo{volume}{14}},
  \bibinfo{pages}{3438} (\bibinfo{year}{1976}).

\bibitem[{\citenamefont{Yu et~al.}(1996)\citenamefont{Yu, Zhu, Zhang, and
  Knobel}}]{fcc}
\bibinfo{author}{\bibfnamefont{R.~H.} \bibnamefont{Yu}},
  \bibinfo{author}{\bibfnamefont{J.}~\bibnamefont{Zhu}},
  \bibinfo{author}{\bibfnamefont{X.}~\bibnamefont{Zhang}}, \bibnamefont{and}
  \bibinfo{author}{\bibfnamefont{M.}~\bibnamefont{Knobel}},
  \bibinfo{journal}{Materials Science and Technology}
  \textbf{\bibinfo{volume}{12}}, \bibinfo{pages}{464} (\bibinfo{year}{1996}).

\bibitem[{\citenamefont{Cullity}(1972)}]{cullity}
\bibinfo{author}{\bibfnamefont{B.~D.} \bibnamefont{Cullity}},
  \emph{\bibinfo{title}{Introduction to magnetic materials}}
  (\bibinfo{publisher}{Addison-Wesley}, \bibinfo{year}{1972}).

\bibitem[{\citenamefont{Yu et~al.}(1995)\citenamefont{Yu, Zhang, Tejada, Zhu,
  Kobel, Tiberto, Allia, and Vinai}}]{yu}
\bibinfo{author}{\bibfnamefont{R.~H.} \bibnamefont{Yu}},
  \bibinfo{author}{\bibfnamefont{X.~X.} \bibnamefont{Zhang}},
  \bibinfo{author}{\bibfnamefont{J.}~\bibnamefont{Tejada}},
  \bibinfo{author}{\bibfnamefont{J.}~\bibnamefont{Zhu}},
  \bibinfo{author}{\bibfnamefont{M.}~\bibnamefont{Kobel}},
  \bibinfo{author}{\bibfnamefont{P.}~\bibnamefont{Tiberto}},
  \bibinfo{author}{\bibfnamefont{P.}~\bibnamefont{Allia}}, \bibnamefont{and}
  \bibinfo{author}{\bibfnamefont{F.}~\bibnamefont{Vinai}}, \bibinfo{journal}{J.
  Appl. Phys.} \textbf{\bibinfo{volume}{78}}, \bibinfo{pages}{5062}
  (\bibinfo{year}{1995}).

\bibitem[{\citenamefont{Bertotti}(1998)}]{bertotti}
\bibinfo{author}{\bibfnamefont{G.}~\bibnamefont{Bertotti}},
  \emph{\bibinfo{title}{Hysteresis in magnetism. For Physicists,materials
  scientists, and engineers}} (\bibinfo{publisher}{Academic Press},
  \bibinfo{year}{1998}).

\bibitem[{\citenamefont{Altbir et~al.}(1996)\citenamefont{Altbir, Vargas, and
  d'~{A}lbuquerque~e Castro}}]{altbir1}
\bibinfo{author}{\bibfnamefont{D.}~\bibnamefont{Altbir}},
  \bibinfo{author}{\bibfnamefont{P.}~\bibnamefont{Vargas}}, \bibnamefont{and}
  \bibinfo{author}{\bibfnamefont{J.}~\bibnamefont{d'~{A}lbuquerque~e Castro}},
  \bibinfo{journal}{Phys. Rev. B} \textbf{\bibinfo{volume}{54}},
  \bibinfo{pages}{R6823} (\bibinfo{year}{1996}).

\bibitem[{\citenamefont{Dormann et~al.}(1997)\citenamefont{Dormann, Fiorani,
  and Tronc}}]{dormann}
\bibinfo{author}{\bibfnamefont{J.~L.} \bibnamefont{Dormann}},
  \bibinfo{author}{\bibfnamefont{D.}~\bibnamefont{Fiorani}}, \bibnamefont{and}
  \bibinfo{author}{\bibfnamefont{E.}~\bibnamefont{Tronc}},
  \bibinfo{journal}{Adv. Chem. Phys.} \textbf{\bibinfo{volume}{XCVIII}},
  \bibinfo{pages}{283} (\bibinfo{year}{1997}).

\end{thebibliography}

\end{document}